\documentclass[twocolumn]{aastex62}

\usepackage{graphicx}
\graphicspath{{./}{figures/}}

\received{}
\revised{}
\accepted{}
\submitjournal{ApJL}

\shorttitle{Lithium-6 in red giants}
\shortauthors{Aguilera-G\'omez et al.}

\begin{document}

\global\csname @topnum\endcsname=1

\title{On Lithium-6 as diagnostic of the lithium-enrichment mechanism in red giants}

\correspondingauthor{Claudia Aguilera-G\'omez}
\email{craguile@uc.cl}

\author[0000-0002-9052-382X]{Claudia Aguilera-G\'omez}
\affil{Departamento de Ciencias F\'isicas Universidad Andr\'es Bello\\ Fernandez Concha 700, 759-1538, Las Condes, Santiago, Chile}

\author[0000-0003-2481-4546]{Julio Chanam\'e}
\affiliation{Instituto de Astrof\'isica, Pontificia Universidad Cat\'olica de Chile\\
Av. Vicu\~na Mackenna 4860, 782-0436 Macul, Santiago, Chile}
\affiliation{Millennium Institute of Astrophysics, Santiago, Chile}

\author[0000-0002-7549-7766]{Marc H. Pinsonneault}
\affiliation{Department of Astronomy The Ohio State University\\
Columbus, OH 43210, USA}



\begin{abstract}
High lithium-7 ($\mathrm{^7Li}$) abundances in giants are indicative of non-standard physical processes affecting the star. Mechanisms that could produce this signature include contamination from an external source, such as planets, or internal production and subsequent mixing to the stellar surface. However, distinguishing between different families of solutions has proven challenging, and there is no current consensus model that explains all the data. The lithium-6 ($\mathrm{^6Li}$) abundance may be a potentially important discriminant, as the relative $\mathrm{^6Li}$ and $\mathrm{^7Li}$ abundances are expected to be different if the enrichment were to come from internal production or from engulfment. In this work, we model the $\mathrm{^6Li}$ and $\mathrm{^7Li}$ abundances of different giants after the engulfment of a substellar mass companion. Given that $\mathrm{^6Li}$ is more strongly affected by Galactic chemical evolution than $\mathrm{^7Li}$, $\mathrm{^6Li}$ is not a good discriminant at low metallicities, where it is expected to be low in both star and planet. For modeled metallicities ([Fe/H]$>-0.5$), we use a ``best case" initial $\mathrm{^6Li/^7Li}$ ratio equal to the solar value.  $\mathrm{^6Li}$ increases significantly after the engulfment of a companion. However, at metallicities close to solar and higher, the $\mathrm{^6Li}$ signal does not last long in the stellar surface. As such, detection of surface $\mathrm{^6Li}$ in metal-rich red giants would most likely indicate the action of a mechanism for $\mathrm{^6Li}$-enrichment other than planet engulfment. At the same time, $\mathrm{^6Li}$ should not be used to reject the hypothesis of engulfment in a $\mathrm{^7Li}$-enriched giant or to support a particular $\mathrm{^7Li}$-enhancement mechanism.

\end{abstract}
 


\section{Introduction} \label{sec:intro}

Lithium-7, one of the two stable isotopes of lithium (Li), was produced right after the Big Bang, and it is used to understand element production in the early Universe \citep{coc14}, diagnose mixing in stellar interiors \citep{pinsonneault97}, and study galactic chemical evolution \citep{prantzos17}, among other applications.

In low-mass stars, $\mathrm{Li}$ is destroyed in the interior during the main sequence. When stars evolve to the red giant branch (RGB), during the first dredge-up the outer convection zone deepens in mass, diluting the $\mathrm{^7Li}$ left close to the stellar surface. For this reason, high $\mathrm{^7Li}$ abundances in giants require the presence of non-standard mechanisms modifying the abundance of the star.

One possible explanation for high $\mathrm{^7Li}$ in the surface of red giants relies on the efficient transport by extra-mixing of $\mathrm{^7Li}$ produced through the Cameron-Fowler mechanism \citep{cameronfowler71}. Another explanation for the enhanced $\mathrm{^7Li}$ is the contamination from a source that preserves or creates $\mathrm{^7Li}$, such as supernovae \citep{martin94} or substellar companions \citep[e.g.,][]{siesslivio99}. An evolved companion, such as an asymptotic giant branch star, which produces $\mathrm{^7Li}$ during its thermal pulses \citep{sackmannboothroyd92} could also be a source of Li. However, the small fraction of Li-rich giants that have been searched for binary companions do not seem to show evidence for them \citep[Chapter 3.1,][]{AG18book}. Further work is needed to test this possibility for the majority of red giants.

In \citet{AG16}, we modeled the engulfment of different planets and brown dwarfs by giant stars. We found that engulfment of substellar companions (SSCs) alone can explain $\mathrm{^7Li}$ abundances as high as $\mathrm{A(^7Li)}=2.2$\footnote{A(x)=$\log (n_x/n_H)+12$}, and that stellar mass and metallicity are fundamental in defining the expected $\mathrm{^7Li}$ abundance in giants and not misinterpret normal giants as enriched, or truly anomalous giants as normal. However, as giants with much higher abundances are found in nature \citep[e.g.][]{Yan18,deepakreddy19}, either a completely different mechanism, or a combination of different $\mathrm{^7Li}$ sources is still needed to explain the entire population.

Other observational indicators can be used to distinguish between different $\mathrm{^7Li}$ replenishment scenarios. The evolutionary phase of the enriched giants is an important indicator of the physical conditions where the enrichment is produced. Some works, such as \citet{deepakreddy19} and \citet{casey19}, argued that most of these unusual giants are located in the horizontal branch. This could point to a mechanism of $\mathrm{^7Li}$ enrichment working during or close to the RGB tip, during the helium flash.
On the other hand, measurements of the stellar rotation \citep{carlberg12}, beryllium surface abundance \citep{takedatajitsu17}, and carbon isotopic ratio \citep{tayar15} could all be fundamental in finding the mechanism behind the $\mathrm{^7Li}$-enrichment.

Another potentially important probe could be $\mathrm{^6Li}$, the far-less-abundant stable isotope of Li, thought to be primarily produced by cosmic ray spallation \citep{meneguzzi71}. 

As $\mathrm{^6Li}$ is destroyed in stellar interiors at even lower temperatures than those required to burn $\mathrm{^7Li}$ \citep{brownschramm88}, standard stellar evolutionary models predict much more severe burning of $\mathrm{^6Li}$ than $\mathrm{^7Li}$ at any evolutionary state \citep{proffittmichaud89}, and very low surface $\mathrm{^6Li}$ abundances during the RGB. 

In contrast, planets and brown dwarfs preserve their initial $\mathrm{^6Li}$, so the abundance of this isotope should be higher in giants that have engulfed their companions. On the contrary, the Cameron-Fowler mechanism is not able to produce $\mathrm{^6Li}$. Thus, it may be possible to use  $\mathrm{^6Li}$ to identify candidates of planet engulfment \citep{charbonnelbalachandran00}.

Because of the large constrast of $\mathrm{^6Li}$ pre and post-engulfment, the planet signal could be easier to detect than that of $\mathrm{^7Li}$. However, at lower metallicities, chemical evolution effects predict very low birth planetary abundances, complicating observations, and the fragility of $\mathrm{^6Li}$ implies that it could be burned even where $\mathrm{^7Li}$ is stable. To test these issues and analyze if $\mathrm{^6Li}$ can effectively be used as a diagnostic of engulfment for all giants,  we model the abundance of $\mathrm{^6Li}$ after the engulfment of SSCs of different properties (Section \ref{sec:models}). The resulting $\mathrm{^6Li}$ surface abundance (Section \ref{sec:li6}) shows that stellar metallicity plays an important role in the burning of $\mathrm{^6Li}$ under convective conditions, with higher metallicity stars burning very rapidly its original $\mathrm{^6Li}$ and that deposited by the planet. As a consequence, the absence of this isotope in the surface of $\mathrm{^7Li}$-rich giants cannot be used to reject the SSC engulfment hypothesis. We analyze in detail this result in Section \ref{sec:disc}, to finally summarize in Section \ref{sec:summary}.

\section{Models} \label{sec:models}
We follow a similar procedure to that described in \citet{AG16}. We refer the reader to that work for an in-depth analysis of the assumptions, the calculation of point of SSC dissipation in stellar interiors, and the parameters used in our grid of stellar models.

In summary, we use a post-processing approach, where standard stellar evolution models are used as a base to later implement the engulfment and thus there is no feedback from the planet ingestion process. Standard stellar models are obtained with the Yale Rotating Evolutionary code \citep{pinsonneault89}.

The modeled stellar mass goes from $1.0$ to $2.0\ \mathrm{M_\odot}$. Metallicities range from [Fe/H]=$-0.5$ up to [Fe/H]=$0.18$ and giants are evolved up to the tip of the RGB. We do not consider lower metallicities because the normal Galactic chemical evolution trends would predict a smaller than solar birth $\mathrm{^6Li/^7Li}$ ratio. In such stars, an engulfed planet is likely to supply little $\mathrm{^6Li}$ due to its low birth $\mathrm{^6Li}$. Thus, the low overall $\mathrm{^6Li}$ would make this signal impossible to observe. Low metallicity stars are also known to experience severe in-situ Li depletion on the giant branch. This combination makes $\mathrm{^6Li}$ a poor discriminant for metal-poor progenitors, and we therefore focus on higher metallicity stars.

The $\mathrm{^6Li}$ in stellar interiors is burned through the reaction
\begin{equation}
\mathrm{^6Li+H\rightarrow {}^3He+{}^4He},
\end{equation}
with reaction rates from \citet{lamia13}.

Regarding the stellar initial abundance of $\mathrm{^6Li}$ in our models, we consider a fixed meteorite Li isotopic ratio $\mathrm{^6Li/^7Li}=0.082$ \citep{chaussidon98}. Because the abundance of $\mathrm{^6Li}$ should increase with metallicity due to the contribution of cosmic ray spallation \citep[e.g.][]{prantzos12}, the birth $\mathrm{^6Li}$ is expected to be lower at lower metallicity.  We therefore regard this as an optimistic or limiting case scenario, where engulfed objects will give the maximum signal. We note, however, that our differential depletion calculations are independent of the assumed birth ratio, given that the $\mathrm{^6Li}$ and $\mathrm{^7Li}$ depletion factors, defined as the fraction of initial Li remaining in the surface of the star, are independent of the birth values.

\begin{figure}
\begin{center}
\includegraphics[width=0.5\textwidth]{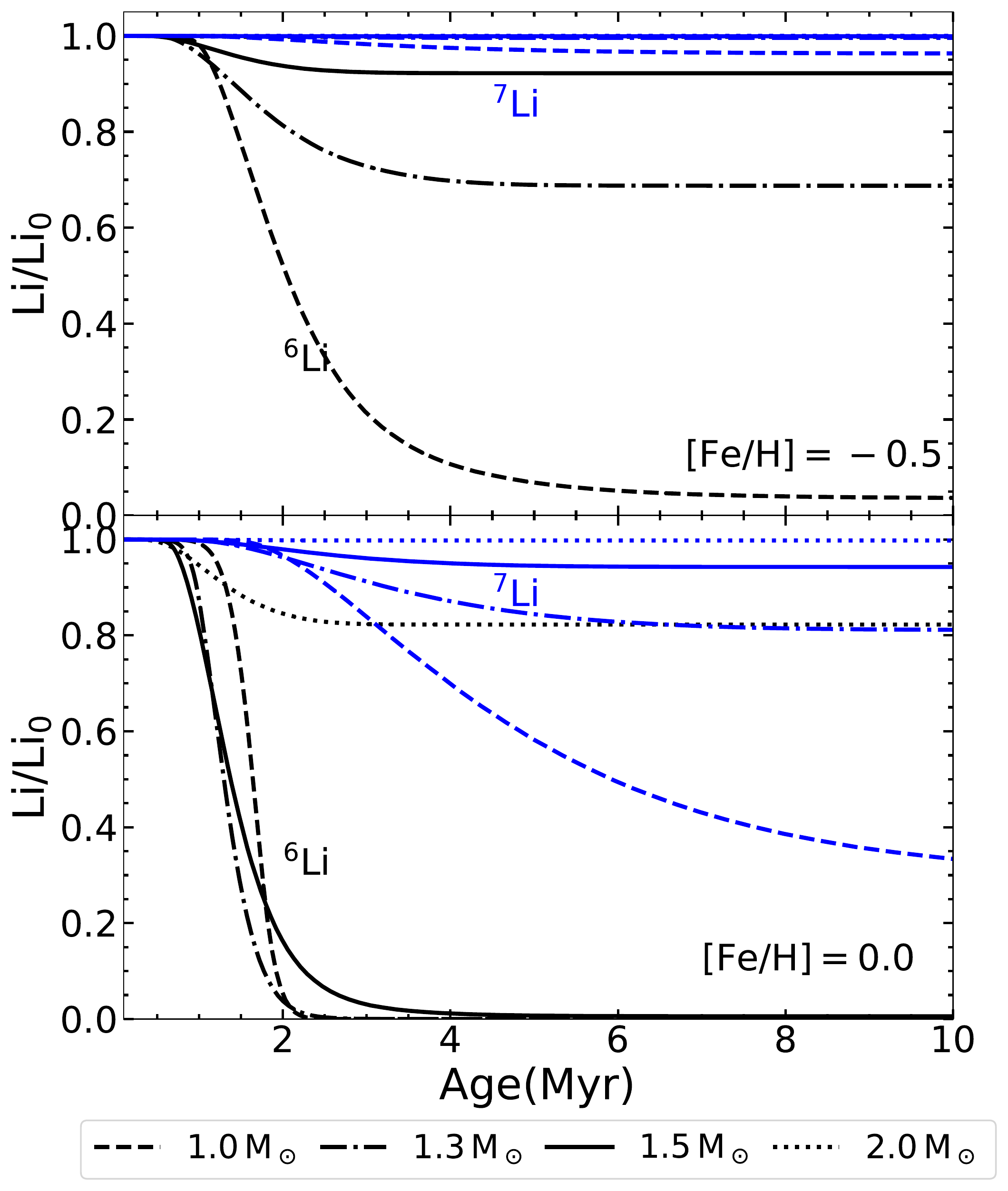}
\caption{$\mathrm{^6Li}$ (black) and $\mathrm{^7Li}$ (blue) in the pre-main sequence of stars of 4 different masses, at metallicities [Fe/H]=$-0.5$ (top panel) and [Fe/H]=$0.0$ (bottom panel).\label{fig:pms}}
\end{center}
\end{figure}

The initial $\mathrm{^6Li}$ value is set before the expected phase of Li burning in the pre-main sequence, thus, the Li isotopic ratio can drastically change in this phase. Figure \ref{fig:pms} shows the burning of Li in the pre-main sequence for stars of different mass and metallicities of [Fe/H]=$-0.5$ (top) and [Fe/H]=$0.0$ (bottom panel). Higher-mass stars preserve their $\mathrm{^6Li/^7Li}$, while there is more burning in solar metallicity stars. 

Notice that the chosen time resolution of the models could change the surface Li abundance in certain models and by using specific settings \citep{lattanzio15}. Here, we test if decreasing the timestep can significantly modify our results, finding that the time resolution only produces slight changes in the abundance.

\begin{figure}
\begin{center}
\includegraphics[width=0.5\textwidth]{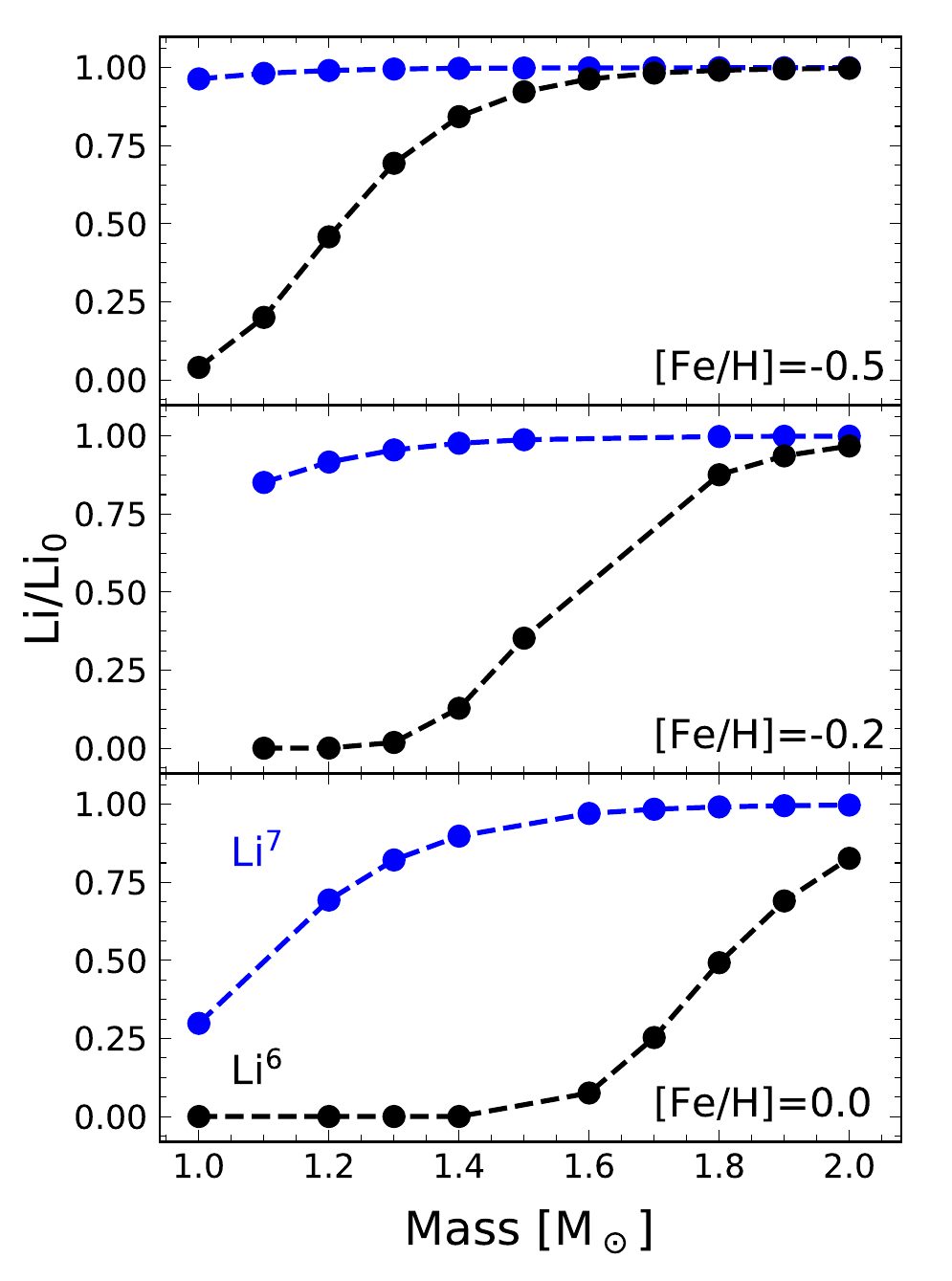}
\caption{$\mathrm{^6Li}$ (black) and $\mathrm{^7Li}$ (blue) depletion factors at the zero-age main sequence (i.e., due to pre-main sequence evolution) for stars of different masses. The panels show results for specific metallicities. \label{fig:zamstams}}
\end{center}
\end{figure}

To better control for the effect of Li burning previous to the RGB phase, we quantify the Li abundances at the zero-age main sequence. Although there is some burning of $\mathrm{^6Li}$ during the main sequence, the main depletion process takes place before that. Figure \ref{fig:zamstams} shows the $\mathrm{^7Li}$ and $\mathrm{^6Li}$ depletion factors at the zero-age main sequence, for stars of different masses and metallicities. There is little to no depletion at higher masses, but important depletion for $\mathrm{^6Li}$ at low masses at any metallicity. $\mathrm{^7Li}$ also burns considerably in low-mass stars at higher metallicities.

For the SSC, we use a fixed ratio between $\mathrm{^6Li}$ mass fraction and metals equal to the Solar System meteoritic value. Thus, all SSCs have the same $\mathrm{X_{^6Li}/Z}$ but could have a different metal content, changing its mass fraction of $\mathrm{^6Li}$. 

The metal content of SSCs depends on their mass. We use three different mass regimes. Brown dwarfs ($15\ \mathrm{M_J}$) can have two different compositions, solar metallicity $Z=Z_\odot$, or brown dwarfs enhanced in metals. Planets ($0.01\,\mathrm{M_J}$ to $15\,\mathrm{M_J}$) are taken to be enhanced in metals as well. Rocky planets (Mass smaller than $0.01\,\mathrm{M_J}$), which include Earth-type objects are considered to have a much higher metal content of $Z=1$.

Results in \citet{AG16} show that very massive brown dwarfs end up dissolving in the radiative interior rather than in the convective envelope. Because of that, we decide to model SSC masses up to $15\ \mathrm{M_J}$. It is important to notice that at higher metallicities the maximum mass of a companion that still dissolves in the convective zone increases \citep[][]{AG16b}.

\section{$\mathrm{^6Li}$ abundance evolution}\label{sec:li6}

We begin by considering the engulfment of four different SSCs by $1.3\ \mathrm{M_\odot}$ and $1.8\ \mathrm{M_\odot}$ red giants of [Fe/H]$=-0.5$, and a $1.7\ \mathrm{M_\odot}$ of [Fe/H]=$0.05$. The companions correspond to a $15\ \mathrm{M_J}$ brown dwarf with $Z=Z_\odot$, a $15\ \mathrm{M_J}$ brown dwarf with $Z=2.5Z_\odot$, a Jupiter-like planet, and an Earth-like planet.

\begin{figure}
\begin{center}
\includegraphics[trim={0cm 1.8cm 2cm 0.8cm}, clip, width=0.4\textwidth]{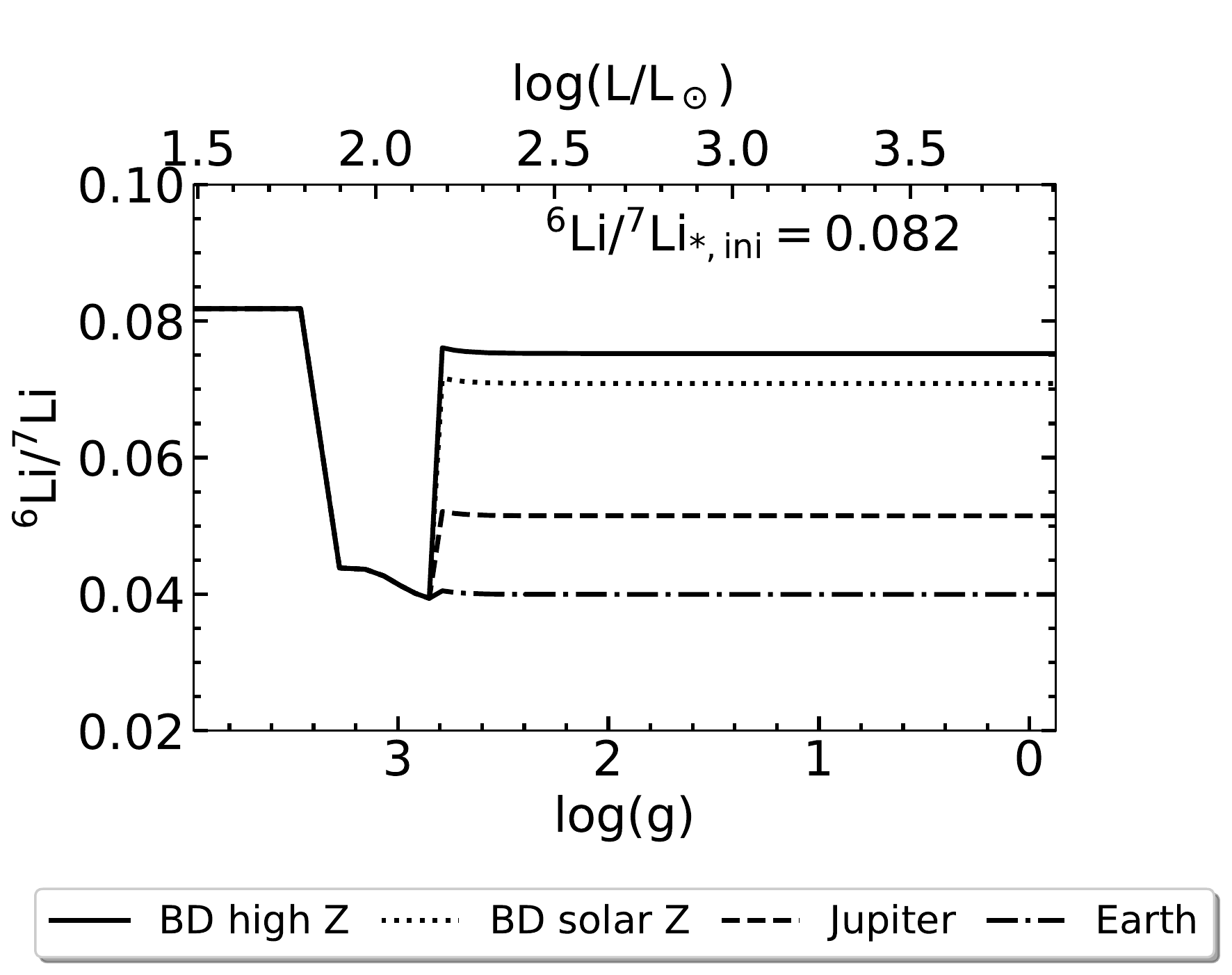}\\
\includegraphics[trim={0cm 1.8cm 2cm 0.8cm},clip, width=0.4\textwidth]{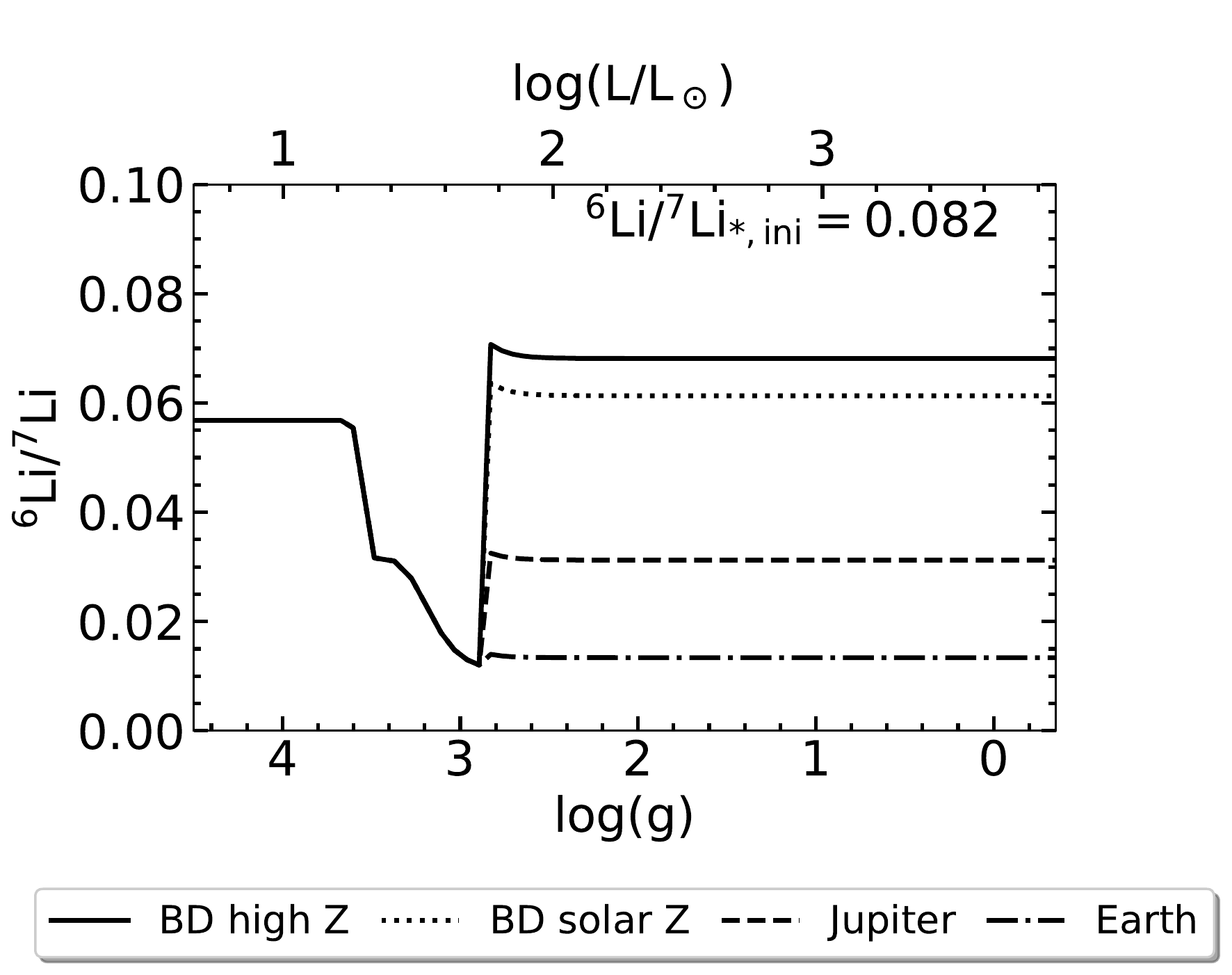}\\
\includegraphics[trim={0cm 1.8cm 2cm 0.8cm},clip, width=0.4\textwidth]{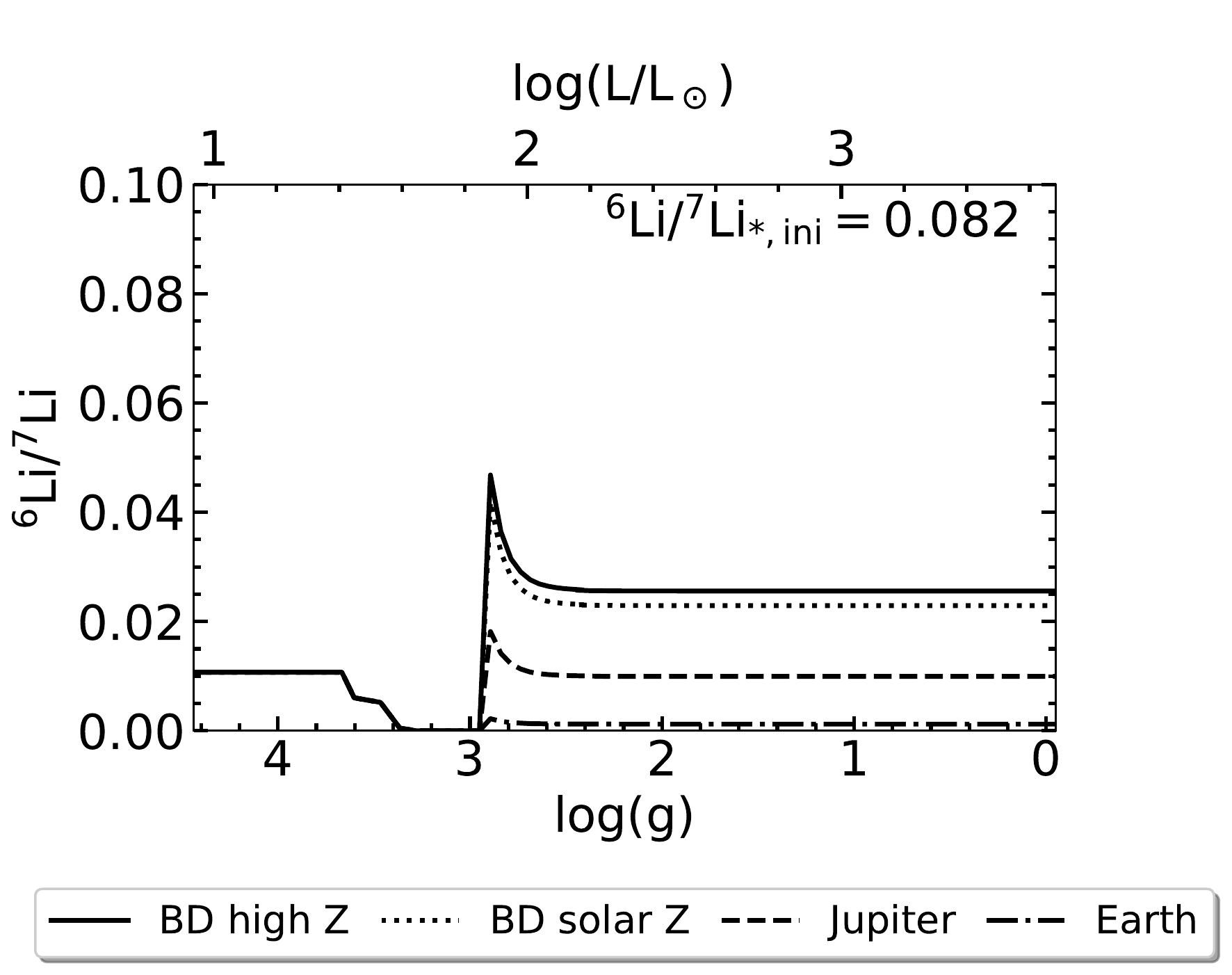}\\
\includegraphics[trim={0cm 0cm 0cm 13cm},clip, width=0.5\textwidth]{m1300z00619ratioli6li7.pdf}
\caption{Surface  $\mathrm{^6Li/^7Li}$ evolution in a $1.8 \ \mathrm{M_\odot}$ star (top panel) and a $1.3\ \mathrm{M_\odot}$ star (middle panel) of [Fe/H]=$-0.5$, and a $1.8\ \mathrm{M_\odot}$ giant of [Fe/H]=$0.05$ (bottom panel) after the engulfment of 4 different SSCs. The evolution starts right before the end of the main sequence and ends at the tip of the RGB. \label{fig:m10}}
\end{center}
\end{figure}

The evolution of the  $\mathrm{^6Li/^7Li}$ surface ratio for these stars can be seen in Figure \ref{fig:m10} as a function of luminosity and $\mathrm{\log g}$. The initial $\mathrm{^6Li}$ in the main sequence can be lower than the meteoritic value due to pre-main sequence burning. The $\mathrm{^6Li/^7Li}$ ratio decreases during the first dredge-up ($\mathrm{\log g}\sim3.5$), as expected. Dilution in the convective envelope decreases the abundance of $\mathrm{^7Li}$ and $\mathrm{^6Li}$. However, the decrease in their ratio is produced because right below the convective envelope, $\mathrm{^6Li}$ burns more rapidly than $\mathrm{^7Li}$. When the first dredge-up mixes that material into the surface, the $\mathrm{^6Li}$ is reduced by a larger amount than $\mathrm{^7Li}$.

The ratio $\mathrm{^6Li/^7Li}$ increase after the engulfment of planets (in our models here, arbitrarily chosen to occur at $\mathrm{\log g}\sim2.8$). The $\mathrm{^6Li}$ enrichment is larger for the brown dwarf with high Z, while Earth-like planets barely increase the original $\mathrm{^6Li}$. 

For giants in the modeled metallicity range, $\mathrm{^6Li}$ burning can be significant during the dredge-up and RGB. We can see this in the $1.7\ \mathrm{M_\odot}$ star in Figure \ref{fig:m10}. Thus, there are some differences in the $\mathrm{^6Li}$ after engulfment in the star when planets are accreated at different locations along the RGB. Later engulfment times imply larger $\mathrm{^6Li}$.

\begin{figure*}
\begin{center}
\includegraphics[width=\textwidth]{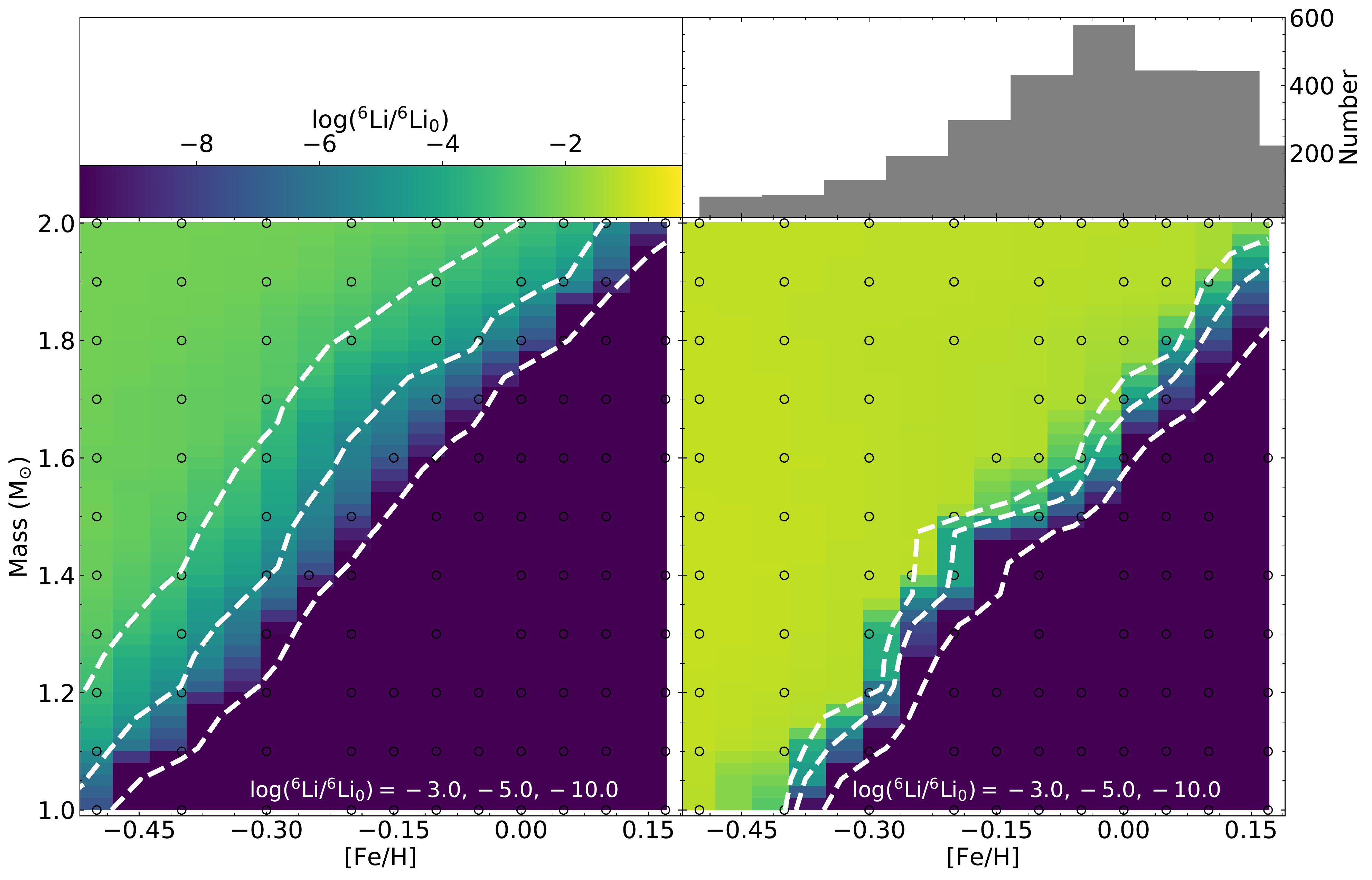}
\caption{Top right panel:  Histogram of the metallicity distribution of giants with measured $\mathrm{^7Li}$, all of them concentrating towards higher metallicities. Bottom left panel: Standard surface $\mathrm{^6Li/^6Li_0}$ abundance of stars of different masses and metallicities. This map considers no engulfment of SSCs. Bottom right panel: Surface $\mathrm{^6Li/^6Li_0}$ abundance of stars of different masses and metallicities after the engulfment of a $15\ \mathrm{M_J}$ brown dwarf enhanced in metal content. In these color maps, grid points are marked with black circles, and the 3 white contours indicate where $\mathrm{log(^6Li/^6Li_0)=}$-3, -5 and -10 from left to right. Stars of higher metallicities burn very rapidly their $\mathrm{^6Li}$ content, as well as any additional $\mathrm{^6Li}$ incorporated by the ingestion of a SSC.   \label{fig:map_npwp}}
\end{center}
\end{figure*}

The resulting $\mathrm{^6Li}$ is mass and metallicity dependent. In Figure \ref{fig:m10}, we see almost no burning post-engulfment in the $1.8 \ \mathrm{M_\odot}$, [Fe/H]$=-0.5$ giant and severe burning in the $1.7 \ \mathrm{M_\odot}$, metal-rich star.

Figure \ref{fig:map_npwp} shows a map of $\mathrm{^6Li/^6Li_0}$ in standard stars of different masses and metallicities, without planet engulfment. We obtain in our models the $\mathrm{^6Li}$ abundance at the tip of the RGB in stars of the grid (small circles in the figure). This grid is then interpolated to produce the map color-coded by $\mathrm{^6Li/^6Li_0}$. For metal-poor stars, a small amount of $\mathrm{^6Li}$ is found in the surface of the star, even without engulfment. However, for metal-rich stars (solar metallicity and higher), the star reaches the RGB with low $\mathrm{^6Li}$, which decreases even more after the first dredge-up. After this stage, $\mathrm{^6Li}$ is also burned under convective conditions, vanishing completely.

Given that the $\mathrm{^6Li}$ is so small in the RGB of standard stars, the engulfment of SSCs could increase substantially the $\mathrm{^6Li}$ abundance. We present a map of $\mathrm{^6Li/^6Li_0}$ for stars of different masses and metallicities in Figure \ref{fig:map_npwp}, bottom right panel, now considering the engulfment of a $15\ \mathrm{M_J}$ brown dwarf enhanced in metals. The giants engulf the SSC at the end of the first dredge-up.  

Comparing this map to the bottom left panel of Figure \ref{fig:map_npwp}, $\mathrm{^6Li}$ can increase significantly with engulfment. However, for metal-rich stars, the incorporated $\mathrm{^6Li}$ is rapidly burned and would not be observed in the stellar surface. This becomes important when distinguishing $\mathrm{^7Li}$-enrichment mechanisms, since most of these giants are metal-rich. We show this in Figure \ref{fig:map_npwp}, top right panel, where we create an histogram of the metallicity of giants with measured $\mathrm{^7Li}$. No upper limits are considered when compiling this catalog, which includes giants from \citet{gilroy89,brown89,jasniewicz99,gonzalez09, kumar11,pace12,carlberg12, lebzelter12, martellshetrone13,liu14,adamow14,bocektopcu15,luck15,carlberg16,DM16,casey16,smiljanic18}; and \citet{deepakreddy19}. These measurements are obtained from the literature, and as such are not homogeneous. Additionally, some of these sources only report their Li-rich giants and not their entire sample\footnote{In \citet{AG16} we find that not reporting the entire sample makes it harder to account for the full phenomenology creating Li-enriched giants.}. As $\mathrm{^7Li}$-rich giants seem to be more metal-rich, this could bias our compilation to higher metallicities.

The limiting metallicity at which $\mathrm{^6Li}$ could never be detected post-engulfment due to its rapid burning increases with mass. For $1.0 \ \mathrm{M_\odot}$, close to [Fe/H]$\sim-0.5$ we already see significant depletion. In $2.0 \ \mathrm{M_\odot}$ giants, this limit is closer to solar metallicity.

If $\mathrm{^6Li}$ is burned in situ the signal of the planet would not be detected. In contrast, the  $\mathrm{^7Li}$ after engulfment could be preserved in the star during the entire RGB phase if no extra-mixing decreases its abundance. This could be the case of more metal-rich stars, where extra-mixing seems to be less-efficient \citep{shetrone19} and indicates that even if the giant accreted a planet, its abundance of $\mathrm{^7Li}$ could be high, while its $\mathrm{^6Li}$ remains low.

\section{Discussion}\label{sec:disc}

As expected, $\mathrm{^6Li}$ can increase in a low-mass red giant after the engulfment of a SSC. However, $\mathrm{^6Li}$ is rapidly burned in stars of higher metallicity, indicating that the absence of this isotope does not discard the possibility that the star has accreted a SSC, but if there was an engulfment event, it did not occur recently. The destruction of this isotope at a faster rate than the $\mathrm{^7Li}$ leads to low $\mathrm{^6Li}$, regardless of the $\mathrm{A(^7Li)}$, not rejecting the engulfment possibility \citep{drake02}. This point therefore becomes a crucial one in the quest for the sources of $\mathrm{^7Li}$ enrichment in giants, as most of the giants that have measured $\mathrm{^7Li}$ have higher metallicites. If $\mathrm{^6Li}$ were to be seen at high metallicity, then its most likely explanation is a source other than an accreted SSC.

At the same time, only the $\mathrm{^7Li}$-rich giants with $\mathrm{A(^7Li)}<2.2$ can be explained by the engulfment of SSC \citep{AG16}. Therefore, the presence or absence of $\mathrm{^6Li}$ in stars of higher $\mathrm{^7Li}$ abundance \citep[e.g.][]{monaco14} does not give any information on this particular enrichment mechanism.

In contrast, if $\mathrm{^6Li}$ is detected in a relatively metal-poor giant with $\mathrm{A(^7Li)}<2.2$, this could be due to the recent contamination of the star by the engulfment of a SSC. Engulfment could explain both the high $\mathrm{^7Li}$ and $\mathrm{^6Li}$ abundances at the same time, but there could also be independent explanations for the enrichment of each isotope.

$\mathrm{^6Li}$ can also be produced in stellar flares \citep{montesramsey98} and galactic cosmic ray interaction with the interstellar medium \citep{fieldsolive99}. Although stellar flares can also produce $\mathrm{^7Li}$, \citet{ramaty00} calculate that the production of the $\mathrm{^6Li}$ isotope is much larger. It is possible that the Sun is producing $\mathrm{^6Li}$ through flares, based on the high abundances found on the lunar soil \citep{chaussidon99}. However, no $\mathrm{^6Li}$ is found in the surface of the Sun, implying that even if some part of the $\mathrm{^6Li}$ created is preserved in the photosphere, it is not enough to be measured. In giants, there is an additional difficulty, given the large convective envelope that would dilute the $\mathrm{^6Li}$ created by any mechanism, complicating its detectability.

From a purely observational point of view, detecting the $\mathrm{^6Li}$ isotope can be particularly hard, as it manifests itself as a subtle asymmetry of the $\mathrm{^7Li}$ line at $\sim6708\,\mathrm{\AA}$. Even a Li isotopic ratio as high as solar can be hard to detect at solar-like metallicites due to convective line asymmetries and blends with other lines. There is a small region of parameter space where the increase in $\mathrm{^6Li}$ could be detected, i.e., in higher mass RGB stars engulfing brown dwarfs companions. These hypothetical detections of $\mathrm{^6Li}$ would be especially interesting in giants with $\mathrm{A(^7Li)}<2.2$. Giants with more $\mathrm{^7Li}$ (and stronger $\mathrm{^7Li}$ lines, where the $\mathrm{^6Li}$ could be more easily detected) can be excluded as engulfment candidates solely based on their $\mathrm{^7Li}$ abundances \citep{AG16}.
However, not only is the $\mathrm{^6Li}$ detection observationally hard, but also, as the stellar mass increases, the lifetime a star spends on its RGB phase decreases considerably. Thus, it is very unlikely to find the higher-mass objects that could retain part of their $\mathrm{^6Li}$ signature.

An interesting solar-metallicity Li-enriched giant is presented by \citet{mott17}, with a $\mathrm{A(^7Li)=1.69}\pm 0.11$ dex. This star has a Li isotopic ratio close to meteoritic. Our models confirm that engulfment is an unlikely explanation for this particular star, that requires further study. 

\section{Summary}\label{sec:summary}

The fragile $\mathrm{^6Li}$ isotope is destroyed at even smaller temperatures than $\mathrm{^7Li}$. As such, stellar evolution theory predicts stars with small $\mathrm{^6Li}$ during the RGB. The $\mathrm{^6Li}$ abundance could increase after the engulfment of SSCs, making $\mathrm{^6Li}$ to appear as a good diagnostic for an engulfment event in giants.

In this work, we found that the $\mathrm{^6Li}$ and $\mathrm{^6Li/^7Li}$ of the star increases after the engulfment of the companion.  We demonstrate that metal-rich stars burn very rapidly the $\mathrm{^6Li}$. The limit between stars that preserve and burn the isotope is mass-dependent.

Given that no $\mathrm{^6Li}$ can be found in metal-rich giants even after planet engulfment, the abundance of this isotope should not be used as a way to distinguish between different $\mathrm{^7Li}$-enrichment mechanisms nor as a method to reject the planet engulfment hypothesis. Moreover, enrichment of $\mathrm{^6Li}$ in low-mass metal-rich giants, is likely not due to planet engulfment. There is only a very low probability that we find such an extremelly recent engulfment event, where $\mathrm{^6Li}$ is still not burned completely.

Stars with $\mathrm{A(^7Li)}>2.2$ could not be explained by planet accretion on the basis of their $\mathrm{^7Li}$ alone. Thus, measurements of $\mathrm{^6Li}$ in these stars do not really indicate anything about the $\mathrm{^7Li}$ enrichment mechanism.
In contrast, finding stars with high abundances of both $\mathrm{^7Li}$ and $\mathrm{^6Li}$ in a certain metallicity range could point to a recent engulfment event. However, a combination of mechanisms, one to enhance $\mathrm{^7Li}$ and another, such as flares, to increase the $\mathrm{^6Li}$, is still possible, especially if the star is metal-rich and its $\mathrm{^6Li}$ is much less likely to be explained by accretion.
In conclusion, we advise caution when using $\mathrm{^6Li}$ as a diagnostic of engulfment or when using it to favor a scenario of $\mathrm{^7Li}$ enrichment over others.

\acknowledgments

We thank G. Somers for his help with lithium in YREC. C.A.G. acknowledges support from the National Agency for Research and Development (ANID) FONDECYT Postdoctoral Fellowship 2018 Project 3180668.  J.C. acknowledges support from CONICYT project Basal AFB-170002 and by the Chilean Ministry for the Economy, Development, and Tourism's Programa Iniciativa Científica Milenio grant IC 120009, awarded to the Millennium Institute of Astrophysics. MHP would like to acknowledge support from NASA grant 80NSSC19K0597.

\bibliographystyle{aasjournal}
\bibliography{Li6_ref}{}

\end{document}